\documentclass[10pt]{article}
\usepackage{euscript,amssymb,epsfig}
\usepackage{amsfonts,latexsym}
\usepackage{euscript,amssymb}
\usepackage{epsfig}

\catcode`\@=11
\def\lesssim{\mathrel{\mathpalette\vereq<}}
\def\vereq#1#2{\lower3pt\vbox{\baselineskip1.5pt \lineskip1.5pt
\ialign{$\m@th#1\hfill##\hfil$\crcr#2\crcr\sim\crcr}}}
\def\gtrsim{\mathrel{\mathpalette\vereq>}}

\def\Let@{\relax\iffalse{\fi\let\\=\cr\iffalse}\fi}
\def\vspace@{\def\vspace##1{\crcr\noalign{\vskip##1\relax}}}
\def\multilimits@{\bgroup\vspace@\Let@
 \baselineskip\fontdimen10 \scriptfont\tw@
 \advance\baselineskip\fontdimen12 \scriptfont\tw@
 \lineskip\thr@@\fontdimen8 \scriptfont\thr@@
 \lineskiplimit\lineskip
 \vbox\bgroup\ialign\bgroup\hfil$\m@th\scriptstyle{##}$\hfil\crcr}
\def\Sb{_\multilimits@}
\def\endSb{\crcr\egroup\egroup\egroup}
\def\Sp{^\multilimits@}

\newcommand{\be}[1]{\begin{equation}\label{#1}}
\newcommand{\ee}{\end{equation}}
\newcommand{\ba}[1]{\begin{eqnarray}\label{#1}}
\newcommand{\ea}{\end{eqnarray}}
\newcommand{\rf}[1]{(\ref{#1})}
\newcommand{\nn}{\nonumber}
\newcommand{\bmatrix}[1]{\left( \begin{array}{#1}}
\newcommand{\ematrix}{\end{array}\right)}

\newlength{\indentedwidth}
\newdimen\mathindent
\indentedwidth=\mathindent

\newenvironment{indented}{\begin{indented}}{\end{indented}}
\newenvironment{varindent}[1]{\begin{varindent}{#1}}{\end{varindent}}
\def\indented{\list{}{\itemsep=0\p@\labelsep=0\p@\itemindent=0\p@
   \labelwidth=0\p@\leftmargin=\mathindent\topsep=0\p@\partopsep=0\p@
   \parsep=0\p@\listparindent=15\p@}\footnotesize\rm}

\def\varindent#1{\setlength{\varind}{#1}%
   \list{}{\itemsep=0\p@\labelsep=0\p@\itemindent=0\p@
   \labelwidth=0\p@\leftmargin=\varind\topsep=0\p@\partopsep=0\p@
   \parsep=0\p@\listparindent=15\p@}\footnotesize\rm}

\begin{document}

\title{Remarks on dimensional reduction of  multidimensional \\ cosmological
models\footnote{Contribution to the proceedings of the Tenth
Marcel Grossmann Meeting on General Relativity, July 20-26, 2003,
Rio de Janeiro, Brazil.}}

\author{Uwe G\"unther$^a$\footnote{e-mail:
u.guenther@fz-rossendorf.de}~\footnote{present address: Research
Center Rossendorf, P.O. Box 510119, D-01314 Dresden, Germany} \
 and Alexander Zhuk$^b$\footnote{e-mail: zhuk@paco.net}\\[2ex] $^a$
Gravitationsprojekt, Mathematische Physik I,\\ Institut f\"ur
Mathematik, Universit\"at Potsdam,\\ Am Neuen Palais 10, PF
601553, D-14415 Potsdam, Germany
\\[1ex] $^b$ Department of Physics, University of Odessa,\\ 2
Dvoryanskaya St., Odessa 65100, Ukraine }

\maketitle

\begin{abstract}Multidimensional cosmological models with
factorizable geometry and their dimensional reduction to effective
four-dimensional theories are analyzed on sensitivity to different
scalings. It is shown that a non-correct gauging of the effective
four-dimensional gravitational constant within the dimensional
reduction results in a non-correct rescaling of the cosmological
constant and the gravexciton/radion masses. The relationship
between the effective gravitational constants of theories with
different dimensions is discussed for setups where the lower
dimensional theory results via dimensional reduction from the
higher dimensional one and where the compactified space components
vary dynamically.
\end{abstract}

\section{Introduction}
\bigskip
One of the basic features of general relativity and of string
theory/M-theory is
 that gravity necessarily propagates in all
dimensions as it inherently describes the dynamics of spacetime
itself. Although this general feature will hold for all
dimensions, its low-energy realization as Gauss's law will
strongly depend on the concrete structure of spacetime, the number
and size of extra-dimensional space components as well as on the
dynamics of the space components and the chosen frame of the
observer. For simplicity, let us consider a multidimensional model
with warped product topology
\be{0.0}
M = M_{D_0}\times M_{D'} \ee
consisting of an external ("our") $D_0-$dimensional spacetime
manifold $M_{D_0}$ (with $D_0=4$) and a $D'$-dimensional compact
space component $M_{D'}$ with characteristic size $L$.
If $r$ is the distance between two massive bodies in a static or
nearly static background metric, then depending on this distance
the masses will attract each other according to Newton's law in
$D$ dimensions for  $r \lesssim L$ and  in $D_0$ dimensions for $r
\gtrsim L$, respectively
\ba{0.01}
F_{D}(r)&=&G_{N(D)}\frac{m_1m_2}{r^{D-2}},\nn\\
F_{D_0}(r)&=&G_{N(D_0)}\frac{m_1m_2}{r^{D_0-2}}.
\ea
The relationship between the fundamental Newton constant
$G_{N(D)}\equiv \kappa_{D}^2/(2 S_{D-1})$ in $D$ dimensions and
the effective Newton constant $G_{N(D_0)}\equiv\kappa_{D_0}^2/(2
S_{D_0-1})$ in the lower dimensional subspace $M_{D_0}$ can be
obtained from Gauss' law in $D$ dimensions \cite{add1,GZ-n1} and
reads
\be{g11}
G_{N(D_0)}=\frac{G_{N(D)}S_{D-1}}{S_{D_0-1}V_{D'}}\, .
\ee
Here, $S_{d}=2\pi ^{d/2}/\Gamma (d/2)$ denotes the surface area of
the unit sphere in $d$ dimensions, and  $V_{D'} \sim L^{D'}$ is
the volume of the extra-dimensional compact space. Because of
$G_{N(4)}\equiv\kappa_{0}^2/(8\pi)=M_{Pl(4)}^{-2}$ and
$G_{N(D)}\equiv \kappa_{D}^2/(2
S_{D-1})=M_{*(4+D^{\prime})}^{-(2+D')}$,\ Eq. \rf{g11} dictates
the relationship \cite{ADD}\footnote{In brane-world models with
non-factorizable geometry, the relation between $\kappa_0$ and
$\kappa_D$ differs from \rf{0.1} \cite{RS,Rubakov}.}
\be{0.1}\kappa_0^2=\kappa_D^2/V_{D'}\quad \Longrightarrow \quad
M_{Pl(4)}^2 \sim V_{D^{\prime}}
M_{*(4+D^{\prime})}^{2+D^{\prime}}\,
\ee
between the Planck scale $M_{Pl(4)} = 1.22\times 10^{19}$GeV and
the fundamental mass scale $M_{*(4+D^{\prime})}$.

 Data from Cavendish-type experiments \cite{exp-1}
confirmed the effective four-dimensionality of our Universe to
high precision at distances above the lower bound of 1 mm. The
upper bound is set by gravity tests in our planetary system,
whereas modifications of Gauss's law at galactic and
inter-galactic scales are not ruled out \cite{DGP-1}. The analysis
of possible observational consequences of extra dimensions at
distances over these well-tested  scales requires not only a
qualitatively correct procedure of dimensional reduction, but also
a quantitatively correct one. The main purpose of this short
remark is to demonstrate this issue, which specifically occurs for
models in an Einstein frame formulation, with the help of a simple
multidimensional cosmological toy model. We also briefly discuss
the frame dependence of the effective Newton's law in the
dimensionally reduced theory and its behavior under a slow
dynamical evolution of the compactification scale of the internal
space components.


To start with, let us consider a cosmological model with
factorizable geometry,
\begin{equation}
\label{2.1}g=g^{(0)}(x)+\sum_{i=1}^n L_{Pl}^2 e^{2\beta
^i(x)}g^{(i)},
\end{equation}
which is defined on the manifold \rf{0.0} where, for generality,
$M_{D'}$ is a direct product of $n$ compact $d_i$-dimensional
spaces: $M_{D'} = \prod\nolimits_{i=1}^n M_i, \;
\sum\nolimits_{i=1}^n d_i = D'$. For simplicity, we assume that
the factors $M_i$ are Einstein spaces: $R_{mn}[g^{(i)}] =
\lambda^i g^{(i)}_{mn},\; m,n = 1,\ldots ,d_i$ and
$R[g^{(i)}]=\lambda^id_i \equiv R_i$. The scale factors of the
internal spaces depend on the coordinates of the four-dimensional
external spacetime, $\beta^i = \beta^i(x)$.

Let $b_i \equiv L_{Pl}e^{\beta^i}$ and $b_{(0)i} \equiv
L_{Pl}e^{\beta^i_0}$ denote the scales factors of the internal
spaces $M_i$ at arbitrary and at present time\footnote{A dynamical
behavior of the extra dimensions results via dimensional reduction
in a variation of the effective physical constants of the
resulting four-dimensional theory (see e.g. section \ref{vary}
below and relation \rf{0.1}, where an uncompensated changing in
time of $V_{D'}$ would lead to the variation of $\kappa_0$). First
discussions of this subject date back to Ref. \cite{horvath}. The
stabilization of internal spaces was discussed, e.g.,  in Refs.
\cite{PRD(1997),PRD(2000)}.}, respectively. Then the total volume
of the internal spaces at the present time is given by
\be{2.7} V_{D^{\prime }}
 \equiv V_I\times v_0 \equiv \prod_{i=1}^n\int\limits_{M_i}d^{d_i}y
\sqrt{|g^{(i)}|} \times \left( \prod_{i=1}^n e^{d_i\beta^i_0}
(L_{Pl})^{D'}\right) = V_I \times \prod_{i=1}^n b_{(0)i}^{d_i}.
\ee
The factor $V_I$ is dimensionless and defined by geometry and
topology of the internal spaces. In this section and the next one,
\be{2.8} \tilde \beta^i = \beta^i - \beta^i_0\,  \ee
denote the deviations of the internal scale factors from their
present day values.

For the demonstration of the scaling sensitivity it is sufficient
to perform the dimensional reduction on the simplest
multidimensional action
\begin{equation} \label{2.6}S=\frac 1{2\kappa_D
^2}\int\limits_Md^Dx\sqrt{|g|}\left\{ R[g]-2\Lambda \right\}
-\frac12 \int\limits_Md^Dx\sqrt{|g|}\left( g^{MN}\partial_M \Phi
\partial_N \Phi + 2 U(\Phi) + \ldots \right)
\end{equation}
for bulk matter in form of a minimally coupled scalar field
$\Phi$. The field $\Phi$ itself can be considered in its zero-mode
approximation. This means that $\Phi$ and $U(\Phi)$ depend only on
the coordinates of the external space, and the dimensional
reduction of the model can be performed by a simple  integration
over the coordinates of the internal spaces.

In the next two sections we consider models with internal space
scale factors which are stabilized in the minimum position of an
effective potential and concentrate on different normalizations of
the gravitational constant $\kappa_0^2$ and the mass scales via
the volume $V_{D'}$. Subject of the third section will be the
frame dependence of changes in the effective Newton's law of the
dimensionally reduced theory when the scale factors are not yet
stabilized. In the last section we briefly summarize our results.
\section{The four-dimensional effective model \label{4d}}

As first step we perform a dimensional reduction of action
\rf{2.6}, what results in the following four-dimensional effective
theory:
\ba{2.9}
S&=&\frac 1{2\kappa _0^2}\int\limits_{M_0}d^{D_0}x\sqrt{|g^{(0)}|}%
\prod_{i=1}^ne^{d_i\tilde \beta ^i}\left\{ R\left[ g^{(0)}\right]
-G_{ij}g^{(0)\mu\nu }\partial _\mu \tilde \beta ^i\,\partial _\nu
\tilde \beta ^j+\right. \nn \\ &+& \left. \sum_{i=1}^n \tilde R_i
e^{-2\tilde \beta^i}-2\Lambda - g^{(0)\mu \nu}\kappa^2_D
\partial_{\mu} \Phi \partial_{\nu} \Phi -2\kappa^2_D
U(\Phi)- \ldots \right\} \, . \ea
Here the total volume of the internal spaces $V_{D'}$ is defined
by Eq. \rf{2.7}, the gravitational constant $\kappa^2_0$ is given
as $\kappa^2_0 = \kappa^2_D/V_{D^{\prime }}$ in accordance with
relation \rf{g11}, and the notations $\tilde R_i := R_i
e^{-2\beta^i_0}L_{Pl}^{-2}$, $G_{ij}=\delta_{ij}d_i-d_i d_j$ are
used. In the ADD approach \cite{add1,ADD} the electro-weak scale
is assumed as the fundamental scale $M_{*(4+D^{\prime})} \sim
M_{EW}$ so that relation \rf{0.1} leads to strong restrictions on
the total volume $V_{D^{\prime }}$ of the internal spaces.

Action \rf{2.9} of the four-dimensional effective model is written
in Brans-Dicke frame, i.e., it has the form of a generalized
Brans-Dicke theory. As next step, we remove the explicit dilatonic
coupling term in \rf{2.9} by conformal transformation
\begin{equation}
\label{2.12} g_{\mu \nu }^{(0)}= \Omega^2 \tilde g_{\mu \nu
}^{(0)} := {\left( \prod_{i=1}^ne^{d_i\tilde \beta ^i}\right) }
^{\frac{-2}{D_0-2}} \tilde g_{\mu \nu }^{(0)}
\end{equation}
and obtain the effective action in the Einstein-frame
\be{2.13}
S=\frac 1{2\kappa _0^2}\int\limits_{M_0}d^{D_0}x\sqrt{|\tilde
g^{(0)}|}\left\{ \tilde R\left[ \tilde g^{(0)}\right] -\bar G_{ij}
\tilde g^{(0)\mu \nu }\partial _\mu \tilde \beta ^i\,
\partial _\nu \tilde \beta ^j - \tilde g^{(0)\mu \nu}\kappa^2_D
\partial_{\mu} \Phi \partial_{\nu} \Phi - 2U_{eff}\right\} ,
\ee
where the effective potential $U_{eff}$ reads
\begin{equation}
\label{2.14}U_{eff}[\tilde \beta ,\Phi ] = {\left(
\prod_{i=1}^ne^{d_i\tilde \beta ^i}\right) }^{-\frac
2{D_0-2}}\left[ -\frac 12\sum_{i=1}^n\tilde R_ie^{-2\tilde \beta
^i}+\Lambda +\kappa_D^2 U(\Phi )+\ldots \right]
\end{equation}
and the notation $\bar G_{ij}=\delta_{ij}d_i+d_i d_j/(D_0-2)$ is
used.
 The internal spaces are stabilized at
present time if this potential has a minimum at $\tilde \beta^i =
0$. Small conformal excitations of the internal spaces above this
minimum have the form of massive scalar fields
(gravexcitons/radions) in our four-dimensional spacetime
\cite{PRD(1997)}. For models with the scalar field $\Phi$ as the
only bulk field, the stabilization occurs for fine-tuned scalar
curvatures of the internal spaces \cite{PRD(2000)}:
\be{2.19} \frac{\tilde R_k}{d_k} = \frac{\tilde R_i}{d_i}\, ,
\quad (i,k = 1,\dots ,n)\, . \ee
For the four-dimensional effective cosmological constant of such
models holds
\be{2.21} \Lambda_{eff} :=\left. U_{eff}\vphantom{\int} \right|_
{\tilde \beta^i =0,\atop \Phi = \Phi_0}\; =\,
\frac{D_0-2}{2}\frac{\tilde R_k}{d_k}\, , \ee
whereas the gravexciton/radion masses are defined by the relations
\be{2.22} m_i^2 = -\frac{4\Lambda_{eff}}{D_0-2}=-2\frac{\tilde
R_k} {d_k} > 0\; ,\quad i = 1,\ldots ,n\, .  \ee
The important point is that the cosmological constant \rf{2.21}
and the gravexciton/radion masses \rf{2.22} in the Einstein frame
are defined at the present time and in a model with present-time
effective gravitational constant $\kappa^2_0 =
\kappa^2_D/V_{D^{\prime }}$. In the next section we will show that
these properties do not hold when the scaling in the dimensional
reduction scheme is chosen differently.
\section{Alternative approach}

In this section, we consider a dimensional reduction scheme which
slightly differs from that of the previous section and which was
employed in many papers. In this scheme the stabilization
positions of the functions $\beta^i$ are not fixed from the very
beginning, but instead they are found from the minimum condition
of the effective potential. In other words, the scale factors
$\beta^i$ are not split into present-time stabilization positions
$\beta_0^i$ and small fluctuational components $\tilde \beta^i$
above them. The dimensional reduction of action \rf{2.6} parallels
that of the previous section with replacing $\tilde \beta^i $
 by $\beta^i$. The result reads
\ba{2.23}
S&=&\frac 1{2\kappa _0^2}\int\limits_{M_0}d^{D_0}x\sqrt{|g^{(0)}|}%
\prod_{i=1}^ne^{d_i \beta ^i}\left\{ R\left[ g^{(0)}\right]
-G_{ij}g^{(0)\mu\nu }\partial _\mu  \beta ^i\,\partial _\nu \beta
^j+\right. \nn \\ &+& \left. \sum_{i=1}^n  R_i e^{-2
\beta^i}-2\Lambda - g^{(0)\mu \nu}\kappa^2_D
\partial_{\mu} \Phi \partial_{\nu} \Phi -2\kappa^2_D
U(\Phi)- \ldots \right\} \, , \ea
with an effective gravitational constant given by
\be{2.24} \kappa^2_0 := \frac{\kappa^2_D}{V_I\times \left(
L_{Pl}\right)^{D'}} \quad \Longrightarrow \quad M_{Pl(4)}^2 \sim
V_I \left( L_{Pl}\right)^{D'} M_{*(4+D^{\prime})}^{2+D^{\prime}}
\, .\ee
In this approach, $V_I \times \left( L_{Pl}\right)^{D'}$ does not
describe the total present-time volume of the internal spaces
(with the exception of the stabilization position $\beta^i=0$),
and, according to the generalized Gauss's law ("gravity propagates
in all dimensions") the constant $\kappa_0$ in Eq. \rf{2.24} does
not correspond to the present-time gravitational constant. Only in
the case of $\beta^i=0$, i.e. for  Planckian stabilization scales
$b_{(0)i}=L_{Pl}$, the constant $\kappa_0^2$ in Eq. \rf{2.24}
correctly corresponds to the four-dimensional Newtonian
gravitational constant. Here, we assumed $V_I \sim
\mathcal{O}(1)$, which usually holds for constant curvature spaces
with normalization $R[g^{(i)}] = \pm d_i(d_i-1)$. The fundamental
mass scale is then of the order of the Planck scale,
$M_{*(4+D^{\prime})} \sim M_{(Pl)4}$.

As next step, we demonstrate how the formal normalization of the
gravitational constant in Eq. \rf{2.24} results in a non-correct
rescaling of the parameters of the four-dimensional effective
theory (in Einstein frame) when $\beta^i_{0} \neq 0$.  The
conformal transformation to the Einstein frame reads:
\be{2.25} g_{\mu \nu }^{(0)}= {\left( \prod_{i=1}^ne^{d_i \beta
^i}\right) } ^{\frac{-2}{D_0-2}} \bar g_{\mu \nu }^{(0)}. \ee
We observe that the four-dimensional Einstein frame metrics for
the natural approach of the previous section (we denote it below
by subscript "1", the corresponding external space metric is
$\tilde g_{\mu \nu }^{(0)}$) and that of the formal approach
 of the present section (subscript "2", external space metric $\bar g_{\mu \nu }^{(0)}$) are connected as
\be{2.26} \bar g_{\mu \nu }^{(0)}= \omega^2 \tilde g_{\mu \nu
}^{(0)}\, ,\quad \omega^2 = \left(
v_0/(L_{Pl})^{D'}\right)^{2/(D_0-2)} = {\left( \prod_{i=1}^ne^{d_i
\beta_0^i}\right) } ^{\frac{2}{D_0-2}}\, .\ee
The Einstein frame form of action \rf{2.23} is
\be{2.27}S=\frac 1{2\kappa
_0^2}\int\limits_{M_0}d^{D_0}x\sqrt{|\bar g^{(0)}|}\left\{ \bar
R\left[ \bar g^{(0)}\right] -\bar G_{ij} \bar g^{(0)\mu \nu
}\partial _\mu  \beta ^i\,
\partial _\nu  \beta ^j - \bar g^{(0)\mu \nu}\kappa^2_D
\partial_{\mu} \Phi \partial_{\nu} \Phi - 2U_{eff}\right\}
\ee
with an effective potential given by
\begin{equation}
\label{2.28}U_{eff}[ \beta ,\Phi ] = {\left( \prod_{i=1}^ne^{d_i
\beta ^i}\right) }^{-\frac 2{D_0-2}}\left[ -\frac 12\sum_{i=1}^n
R_ie^{-2 \beta ^i}+\Lambda +\kappa_D^2 U(\Phi )+\ldots \right]\, .
\end{equation}
The gravexciton/radion masses as well as the effective
cosmological constants in the two approaches are connected by the
following rescaling,
\ba{2.29} \left. m^2_i \right|_{2}&= &  \left(\prod_{i=1}^n
e^{d_i\beta_0^i}\right)^ \frac{-2}{D_0-2} \left. m_i^2 \right|_{1}
\, , \nn \\ &\phantom{-}& \\ \left. \Lambda_{eff} \right|_{2}&= &
\left(\prod_{i=1}^ne^{d_i\beta_0^i}\right)^ \frac{-2}{D_0-2}
\left. \Lambda_{eff}\right|_{1}
\, , \nn \ea which results from the rescaled effective potential
\be{2.30} \left. U_{eff} \right|_{2}= \left.
\left(\prod_{i=1}^ne^{d_i\beta_0^i}\right)^ \frac{-2}{D_0-2}
U_{eff}\right|_{1} = \left. \omega^{-2}\; U_{eff}\right|_{1}\,
.\ee
In its turn, the rescaling of the effective potential is caused by
the formal and non-correct definition \rf{2.24} of the
four-dimensional gravitational constant $\kappa_0$ in the second
approach. The gravitational constants in the two approaches are
connected by the relation
\be{2.31} \left. \kappa^2_0\right|_{2} = \left.
\prod_{i=1}^ne^{d_i \beta_0^i}\; \kappa^2_0 \right|_{1} = \left.
\omega^{D_0-2}\; \kappa^2_0\right|_{1}\, , \ee
which can also be obtained from the equation
\be{2.32} \frac{1}{\left. 2\kappa^2_0\right|_{2}}\int d^{D_0}x
\sqrt{|\bar g^{(0)}| }\bar R [\bar g^{(0)}] =
\frac{\omega^{D_0-2}}{\left. 2\kappa^2_0\right|_{2}}\int d^{D_0}x
\sqrt{|\tilde g^{(0)}| }\tilde R [\tilde g^{(0)}] \equiv
\frac{1}{\left. 2\kappa^2_0\right|_{1}}\int d^{D_0}x \sqrt{|\tilde
g^{(0)}| }\tilde R [\tilde g^{(0)}]\, .\ee
Thus, if we restore in the second approach the correct
four-dimensional gravitational constant, we obtain the same
formulas as in the first approach:
\ba{2.33}\left.
S\right|_{2}&=&\frac{\omega^{D_0-2}}{\omega^{D_0-2}\left. 2\kappa
_0^2\right|_{2}}\int\limits_{M_0}d^{D_0}x\sqrt{|\bar
g^{(0)}|}\left\{ \bar R\left[ \bar g^{(0)}\right] -\bar G_{ij}
\bar g^{(0)\mu \nu }\partial _\mu  \beta ^i\,
\partial _\nu  \beta ^j - \bar g^{(0)\mu \nu}\kappa^2_D
\partial_{\mu} \Phi \partial_{\nu} \Phi - \left. 2U_{eff}\right|_{2}\right\}\nn \\
&=&\frac 1{\left. 2\kappa
_0^2\right|_{1}}\int\limits_{M_0}d^{D_0}x\omega^{-D_0}\sqrt{|\bar
g^{(0)}|}\left\{ \omega^2 \bar R\left[ \bar g^{(0)}\right]
-\omega^2 \bar G_{ij} \bar g^{(0)\mu \nu }\partial _\mu  \beta
^i\,
\partial _\nu  \beta ^j - \omega^2 \bar g^{(0)\mu \nu}\kappa^2_D
\partial_{\mu} \Phi \partial_{\nu} \Phi - \left. 2\omega^2 U_{eff}\right|_{2}
\right\}\nn \\ &=&\frac 1{\left. 2\kappa
_0^2\right|_{1}}\int\limits_{M_0}d^{D_0}x\sqrt{|\tilde
g^{(0)}|}\left\{ \tilde R\left[ \tilde g^{(0)}\right] -\bar G_{ij}
\tilde g^{(0)\mu \nu }\partial _\mu \tilde \beta ^i\,
\partial _\nu \tilde \beta ^j - \tilde g^{(0)\mu \nu}\kappa^2_D
\partial_{\mu} \Phi \partial_{\nu} \Phi -\left. 2U_{eff}\right|_{1}\right\}
\nn\\ & \equiv & \left. S\right|_{1}\, . \ea

\section{Frame dependence of Newton's gravitational force law
and a possible dynamics of the gravitational
"constant"\label{vary}} In the previous sections the main emphasis
was laid on the correct scaling of the gravitational constant in
present-time physical regimes when the sizes of the internal
spaces are stabilized, and only small fluctuations over this
stabilized scale factor background remain. In general, this
stabilized scale factor regime will be the result and current end
point of highly dynamical changes of the sizes of the internal
spaces at early stages of the cosmological evolution before
nucleosynthesis started\footnote{Observation data strongly
restrict late-time variations of coupling constants, such as the
fine structure constant $\alpha$ and  gravitational constant
$\kappa_0$ \cite{uzan}.}. Taking into account that the effective
gravitational constant $\kappa_0^2$ of our four-dimensional
external spacetime is a derived constant which follows via
dimensional reduction from a fundamental mass scale, we are
naturally led to the question, whether $\kappa_0^2$ also varies
dynamically with the sizes of the internal space components and
correspondingly depends on the external coordinates
$\kappa_0=\kappa_0(x)$, or whether it remains fixed by some
mechanism. Below we will show that the answer of this question
depends on the chosen metric frame of the effective
four-dimensional spacetime.

We start the consideration by splitting the internal space scale
factors
\be{3.1}
\beta^i(x)=\beta^i_0(x)+\tilde{\beta}^i(x)
\ee
into a slowly and coherently changing background component
$\beta^i_0(x)$, which will define the averaged dynamics of the
volume of the internal spaces, and small non-coherent
particle-like excitations/fluctuations $\tilde{\beta}^i(x)$ over
this background, which as previously will correspond to
gravexcitons/radions. The dynamically averaged volume $V_{D'}(x)$
of the internal spaces is then defined in analogy with \rf{2.7} as
\ba{3.2}
 V_{D^{\prime }}(x)
 \equiv V_I\times v_0(x)& \equiv & \prod_{i=1}^n\int\limits_{M_i}d^{d_i}y
\sqrt{|g^{(i)}|} \times \left( \prod_{i=1}^n e^{d_i\beta^i_0(x)}
(L_{Pl})^{D'}\right)\nn \\ &=& V_I \times \prod_{i=1}^n
b_{(0)i}^{d_i}(x).
\ea
Subsequently, we present a sketchy non-relativistic analysis in
terms of Newton's force law. We assume that the scale factor
background changes slowly and smoothly enough, and the
gravexciton/radion amplitudes are small enough to keep inside this
approximation.

Next, we note that the effective gravitational force
\be{3.2a}
F_{eff}(r) =\frac{G_{N(D)}S_{D-1}}{S_{D_0-1}V_{D'}}\frac{m_1
m_2}{r^{D_0 -2}}
\ee
between two masses $m_1$ and $m_2$ separated at a distance $r\gg
L\sim V_{D'}^{1/D'}$ in the external spacetime $M_{D_0}$ is the
result of a dimensional reduction\footnote{For the technical
details we refer to \cite{GZ-n1}.} performed in the starting
metric \rf{2.1} of the total product space $M_D$. The distance $r$
is correspondingly measured in the Brans-Dicke metric $g^{(0)}$ of
the external spacetime $M_{D_0}$ and $V_{D'}$ is the total volume
of the internal space. This means that formally the constant
volume
 $V_{D'}$ of a static internal space, for which the force law was derived, should be replaced by the
total volume $V_{D',T}$ which is now defined by the non-truncated
internal space scale factors $\beta^i(x)$:
\ba{3.2a1}
V_{D'}\mapsto V_{D',T}(x)&=&V_I\times L_{Pl}^{D'}\prod_{i=1}^n e^{d_i\beta^i(x)}\nn\\
&=& V_{D'}(x)\times \prod_{i=1}^n e^{d_i\tilde \beta^i(x)}.
\ea
It is clear that for large non-adiabatic and particle-like scale
factor fluctuations $\tilde \beta^i(x)$ in \rf{3.1} the force law
approximation \rf{3.2a} will break down and should be replaced by
a field theoretic treatment based on a Green function technique as
it was discussed, e.g., in Ref. \cite{add1}. Subsequently, we will
restrict our attention to two regimes for which the Newton's law
\rf{3.2a} with the replacement \rf{3.2a1} can be used as rough
approximation of the gravitational attraction force: a long
wave-length regime (I) without small short wave-length
contributions so that a splitting \rf{3.1} is not
required\footnote{This means that the total internal space volume
$V_{D',T}(x)$ varies slowly enough
 that the gravitational force law will dominate over
gravitational wave effects due to changes of the metric.}
$[\beta^i(x)=\beta^i_0(x)]$; and a regime where sufficiently small
gravexciton/radion fluctuations $\tilde
\beta^i(x)$ over a coherently varying long-wavelength background are present
(regime II).

We will now analyze how the variations of $V_{D',T}(x)$ will
affect the force law \rf{3.2a} in different frames of the external
spacetime $M_{D_0}$. (We assume $D_0=4$.)

\subsection{Brans-Dicke frame}

In Netwon's force law \rf{3.2a} the distance $r$ is measured in
the metric $g^{(0)}$ which coincides with the Brans-Dicke (BD)
frame metric. This means that, in the BD frame, the only change in
the force law \rf{3.2a} will be induced by the replacement of the
total internal space volume according to \rf{3.2a1}:
\be{3.2b}
F_{eff}(r) =\frac{G_{N(D)}S_{D-1}}{S_{D_0-1}V_{D',T}(x)}\frac{m_1
m_2}{r^{D_0 -2}}.
\ee
As result the varying total internal space volume $V_{D',T}(x)$
leads to corresponding variations in the gravitational force
between the masses $m_1$ and $m_2$ in the external space
$M_{(D_0=4)}$. In both regimes we can define a varying effective
gravitational "constant" as
\be{3.3}
\kappa_0^2(x)=\frac{\kappa_D^2}{V_{D',T}(x)}=\frac{\kappa_D^2}{
V_I (L_{Pl})^{D'}}\prod_{i=1}^n e^{-d_i\beta^i_0(x)}.
\ee
For absent scale factor fluctuations (regime I) $\beta^i_0(x)$
coincides with $\beta^i(x)$, whereas in regime II the small
gravexciton/radion-like scale factor fluctuations $\tilde
\beta^i(x)$ are split off from the slowly varying volume.
In the latter regime (II) the action functional \rf{2.9}
reads
\ba{3.4-II}
S&=&\int\limits_{M_0}d^{D_0}x\sqrt{|g^{(0)}|}\frac 1{2\kappa
_0^2(x)} \prod_{i=1}^ne^{d_i\tilde \beta ^i}\left\{ R\left[
g^{(0)}\right] -G_{ij}g^{(0)\mu\nu }\partial _\mu
\beta_0^i\,\partial _\nu  \beta_0^j-2G_{ij}g^{(0)\mu\nu }\partial
_\mu \beta_0 ^i\,\partial _\nu \tilde \beta ^j- \right. \nn \\ &&
- G_{ij}g^{(0)\mu\nu }\partial _\mu \tilde \beta ^i\,\partial _\nu
\tilde \beta ^j+ \left. \sum_{i=1}^n \tilde R_i e^{-2\tilde
\beta^i}-2\Lambda - g^{(0)\mu \nu}\kappa^2_D
\partial_{\mu} \Phi \partial_{\nu} \Phi -2\kappa^2_D
U(\Phi)- \ldots \right\}
\ea
and transition to regime I consists simply in the substitution
$\tilde
\beta^i(x)\to 0$, $\beta^i_0(x) \to \beta^i(x)$.

 In  action functional \rf{3.4-II} the
background scale factors $\beta^i_0(x)$ are via internal space
volume $V_{D'}(x)$ and relation \rf{3.3} responsible for the
variations of the effective gravitational "constant"
$\kappa_0^2(x)$, whereas the small scale factor fluctuations
$\tilde
\beta^i(x)$ can be considered as scalar particles
(gravexcitons/radions) propagating over this background in the
spacetime $M_{D_0}$.

\subsection{Hybrid frame}
Removing the Brans-Dicke factor of the fluctuational components in
action \rf{3.4-II} by a conformal transformation of type \rf{2.12}
\be{3.4a}
g_{\mu \nu }^{(0)}= {\left( \prod_{i=1}^ne^{d_i\tilde \beta
^i}\right) } ^{\frac{-2}{D_0-2}} \hat{g}_{\mu \nu }^{(0)},
\ee
we arrive at an action functional
\ba{3.5}
S&=&\int\limits_{M_0}d^{D_0}x\sqrt{\left|\hat{g}^{(0)}\right|}\frac
1{2\kappa _0^2(x)}\left\{ \tilde R\left[
\hat{g}^{(0)}\right]-G_{ij}\hat{g}^{(0)\mu\nu }\partial _\mu
\beta_0^i\,\partial _\nu  \beta_0^j-2G_{ij}\hat{g}^{(0)\mu\nu }\partial
_\mu \beta_0 ^i\,\partial _\nu \tilde \beta ^j-\right. \nn \\ && -
\bar G_{ij} \hat{g}^{(0)\mu \nu }\partial _\mu \tilde \beta ^i\,
\partial _\nu \tilde \beta ^j -\left. \hat{g}^{(0)\mu \nu}\kappa^2_D
\partial_{\mu} \Phi \partial_{\nu} \Phi - 2U_{eff}\right\} ,
\ea
which is in a hybrid Brans-Dicke-Einstein frame --- Brans-Dicke
frame with respect to the averaged scale factors $\beta_0^i(x) $,
which define the effective gravitational constant $\kappa_0(x)$
via \rf{3.3}, and Einstein frame with respect to the
gravexcitons/radions $\tilde \beta^i (x)$.

For a stabilization of the internal spaces with volume freezing we
have
\be{3.6a}
V_{D^{\prime }}(x)\longrightarrow V_{D^{\prime }}, \qquad
\beta_0^i(x)\longrightarrow \beta_0^i, \qquad
\kappa_0(x)\longrightarrow \kappa_0
\ee
and we return to the simplified action functional \rf{2.13} of
section \ref{4d}. We note, that similarly like possible
present-time variations of the fine-structure "constant" $\alpha $
are strongly restricted by observational data \cite{uzan,GSZ},
there exist strong observational restrictions on present-time
variations of the effective gravitational constant $\kappa_0$
\cite{uzan,kubyshin}\footnote{For completeness, we note that in
string-theoretic setups an additional dynamically changing dilaton
factor enters the definition \rf{3.3} of the effective
four-dimensional gravitational "constant" $\kappa_0$. (See, e.g.,
Ref. \cite{uzan}.)}.

Let us now analyze how the starting Newton's gravitational force
\rf{3.2a} should be changed for an observer in the hybrid frame.
{}From the conformal transformation \rf{3.4a} follows the relation
\be{3.6b}
r={\left( \prod_{i=1}^ne^{d_i\tilde \beta ^i(x)}\right) }
^{\frac{-1}{D_0-2}}\hat r
\ee
between the distances $r$ and $\hat r$ in BD and hybrid frame. On
the other hand, the exact relation for the total internal space
volume reads
\be{3.6c}
V_{D',T}(x)=V_{D'}(x)\times {\left( \prod_{i=1}^ne^{d_i\tilde
\beta (x) ^i}\right) }.
\ee
Plugging Eqs. \rf{3.6b}, \rf{3.6c} into \rf{3.2b} we obtain the
force law in the hybrid frame as
\be{3.6d}
F_{eff}(r) =\frac{G_{N(D)}S_{D-1}}{S_{D_0-1}V_{D'}(x)}\frac{m_1
m_2}{\hat r^{D_0 -2}}\quad \Longrightarrow \quad
\kappa_0^2(x)=\kappa_D^2/V_{D'}(x)\, ,
\ee
i.e., the contributions of the small scale factor fluctuations
$\tilde
\beta ^i(x)$ from total volume and the transformed distance cancelled each other and we are left
with a Newton's force law which depends only on the slowly varying
part $V_{D'}(x)$ of the internal space volume. This means that for
an observer in hybrid frame there is no need of averaging over
small scale factor fluctuations because they are formally
cancelling in the force law.

\subsection{Einstein frame}
Similar to the Einstein frame setup of section \ref{4d}, we assume
the internal space scale factors split into constant background
components $\beta^i_0$ and non-constant components $\tilde
\beta^i(x)$
\be{3.7}
\beta^i(x)=\beta^i_0+\tilde
\beta^i(x).
\ee
The total internal space volume is then defined as
\be{3.8}
V_{D',T}(x)=V_{D'}\times {\left( \prod_{i=1}^ne^{d_i\tilde
\beta (x) ^i}\right) }\, ,
\ee
with $V_{D'}$ given in Eq. \rf{2.7}. We note that, in general, the
constant volume $V_{D'}$  can be interpreted as an arbitrarily
fixed reference volume. For simplicity, we assume in our heuristic
considerations that the non-constant scale factor components
$\tilde \beta^i(x)$ are either slowly varying or, e.g., for
stabilized scale factor backgrounds, sufficiently small to keep
the description in terms of Newton's law physically sensible.

Along the same scheme as for the hybrid frame, we obtain from the
conformal transformation \rf{2.12} a relation between the distance
$r$ measured in the BD frame and the corresponding distance $r_E$
in the Einstein frame
\be{3.9}
r= {\left( \prod_{i=1}^ne^{d_i\tilde \beta ^i}\right) }
^{\frac{-1}{D_0-2}}r_E.
\ee
Substitution of Eqs. \rf{3.8}, \rf{3.9} into \rf{3.2b} yields
\be{3.10}
F_{eff}(r) =\frac{G_{N(D)}S_{D-1}}{S_{D_0-1}V_{D'}}\frac{m_1
m_2}{r_E^{D_0 -2}}\quad \Longrightarrow \quad
\kappa_0^2=\kappa_D^2/V_{D'}.
\ee
Again the contributions of the non-constant scale factor
components $\tilde \beta^i(x)$ cancelled. We observe the
interesting fact that in the Einstein frame the effective Newton's
law in the dimensionally reduced theory does not change,
irrespective of the dynamically changing volume $V_{D',T}(x)$ of
the internal spaces, i.e., the changes in $V_{D',T}(x)$ are
exactly compensated by inverse changes of the Einstein frame
metric $\bar g^{(0)}$ with respect to the original external space
metric $ g^{(0)}$. Hence, in the chosen oversimplified model an
observer in Einstein frame will not be aware of the internal space
dynamics in measurements based on non-relativistic approximations
of the gravitational sector. This is also visible from the
Einstein frame action \rf{2.13}, where the $D_0-$dimensional
gravity sector is given by a pure Einstein-Hilbert term with
non-varying gravitational constant $\kappa_0^2$. For given mass
scales $M_{Pl}=\sqrt{8\pi}/ \kappa_0$,
$M_{*(4+D')}=(2S_{D-1}/\kappa_D^2)^{1/(2+D')}$ relation \rf{3.10}
fixes the internal space (reference) volume $V_{D'}$ completely.

Finally, we note that in the special case of a solely time
dependent scale factor dynamics, the metric structure roughly
parallels  the M-theory inspired cosmological toy models of Ref.
\cite{SUGRA} which live in warped product metrics of the type
\be{m7}
g={\left( \prod_{i=1}^ne^{d_i\tilde \beta ^i(t)}\right) }
^{\frac{-2}{D_0-2}} \tilde g_{\mu \nu }^{(0)}+\sum_{i=1}^n
L_{Pl}^2 e^{2\beta ^i(t)}g^{(i)}.
\ee
\section{Conclusions}

An essential part of any viable higher dimensional theory is a
sensible scheme of dimensional reduction to an effective
four-dimensional theory. On the one hand, the resulting effective
theory should correctly describe our observable four-dimensional
Universe (the external spacetime). On the other hand, it will
contain explicit imprints and signatures of the extra-dimensional
space components. The latter will give an opportunity to predict
new observable phenomena, such as gravexcitons/radions which are
geometrical moduli excitations of extra-dimensional spaces
propagating as special types of particles in the observable
Universe. A correct quantitative prediction of the derived
physical parameters (such as the effective "fundamental"
constants, the cosmological constant, the masses of
gravexcitons/radions etc.) directly depends on the concrete scheme
of dimensional reduction. In the present paper, we demonstrated
this fact explicitly with the help of a multidimensional toy model
with factorizable geometry as it is often used in KK and ADD
approaches. For a model with stabilized internal spaces, we
considered two different schemes of dimensional reduction with
subsequent transformation to the Einstein frame. In the second
approach, the present-time size of the internal spaces was not
taken into account, and, correspondingly, it was left out of
account that gravity should propagate in all dimensions. As a
result, the effective four-dimensional gravitational constant was
not correctly gauged what finally led to a non-correct rescaling
of the parameters of the effective four-dimensional model (such as
the effective cosmological constant and the gravexciton/radion
masses).

Additionally, we discussed the relation between the chosen
observer frame in the external spacetime $M_{D_0}$ and the
dependence of the effective gravitational coupling "constant"
$\kappa_0^2$ in $M_{D_0}$ on the dynamics of the compactified
internal factor space. It was shown that in Brans-Dicke frame and
a hybrid Brans-Dicke-Einstein frame $\kappa_0^2$ changes
dynamically when the volume $V_{D'}(x)$ of the internal space
changes, whereas in Einstein frame $\kappa_0^2$ can be hold fixed
independently of the scale factor dynamics of the internal space.
This constancy of $\kappa_0^2$ results from the special tuning
between the internal space scale factors and the conformal factor
of the external spacetime.

\vspace*{1ex}

\mbox{} \\ {\bf Acknowledgments}\\ UG acknowledges financial
support from DFG grant KON/1344/2001/GU/522. AZ thanks the members
of the Organizing Committee of MG10 for their kind invitation to
present a talk and for their financial support.



\end{document}